\documentstyle[aps,prb,amssymb,epsfig]{revtex}
\begin{document}

\twocolumn[
\hsize\textwidth\columnwidth\hsize
\csname@twocolumnfalse\endcsname

\draft
\preprint{SNUTP 99-020}
\title{Charge Frustration Effects in Capacitively Coupled
 Two-Dimensional Josephson-Junction Arrays}
\author{Minchul Lee, Mahn-Soo Choi$^*$, and M.~Y. Choi}
\address{Department of Physics and Center for Theoretical Physics,
 Seoul National University, Seoul 151-742, Korea}
\maketitle

\begin{abstract}
 We investigate the quantum phase transitions in two capacitively coupled
 two-dimensional Josephson-junction arrays with charge frustration. 
 The system is mapped onto the $S=1$ and $S=1/2$ anisotropic Heisenberg
 antiferromagnets near the particle-hole symmetry line
 and near the maximal-frustration line, respectively, which are in turn
 argued to be effectively described by a single quantum phase model.
 Based on the resulting model, it is suggested that near the
 maximal frustration line the system may undergo a quantum phase transition
 from the charge-density wave to the super-solid phase, which displays both
 diagonal and off-diagonal long-range order.
\end{abstract}
\pacs{PACS numbers: 74.50.+r, 67.40.Db, 73.23.Hk}

] 

%
\newcommand\beq{\begin{equation}}
\newcommand\eeq{\end{equation}}
\newcommand\beqz{\[}
\newcommand\eeqz{\]}
\newcommand\beqa{\begin{eqnarray}}
\newcommand\eeqa{\end{eqnarray}}
\newcommand\beqaz{\begin{eqnarray*}}
\newcommand\eeqaz{\end{eqnarray*}}
\newcommand\bsplit{\begin{eqnarray}}
\newcommand\esplit{\end{eqnarray}}
\newcommand\balign{\begin{eqnarray}}
\newcommand\ealign{\end{eqnarray}}
\newcommand\half{\frac{1}{2}}
\newcommand\lavg{\left\langle\:}
\newcommand\ravg{\:\right\rangle}
\newcommand\bfr{{\bf r}}
\newcommand\bfe{{\bf e}}
\newcommand\BfC{{\mathbb{C}}}
\newcommand\varE{{\cal E}}

%

In recent years, various types of cotunneling transport have been of great
interest in ultrasmall tunnel junctions, which exhibit strong Coulomb
blockade effect.\cite{Averin92}  In particular, cotunneling of the
electron-hole pairs in two capacitively coupled one-dimensional (1D) arrays
of small metallic junctions has been proposed
theoretically~\cite{Averin91a} and demonstrated
experimentally,~\cite{Matter97} revealing the remarkable effects of the current
mirror.  More recently, in capacitively coupled
1D~\cite{ChoiMS98f,ChoiMS97p7} or two-dimensional~\cite{ChoiMS98e}
(2D) Josephson-junction arrays (JJAs), the cotunneling of
particle-hole pairs (with the {\em particle\/} and {\em hole\/} standing for the
{\em excess\/} and {\em deficit\/} Cooper pair, respectively)
has been proposed even to
drive the quantum phase transition from superconductor to insulator (SI) at
zero temperature.  Here the particle-hole symmetry of the system may be broken
by, e.g., the gate voltage applied between the
array and the substrate. The resulting charge frustration is expected to affect
the phase transition of the system in a crucial way. For example, when the
particle-hole symmetry is broken maximally, the transport is governed by
the cotunneling of the particle-void pairs (with the {\em void} denoting
the absence of an excess or deficit Cooper pair) and the different nature of
the associated phase transition has been pointed out in one dimension.\cite{ChoiMS98f}  
On the other hand, existing studies of coupled 2D arrays with charge frustration
have concentrated upon the charge-vortex duality,\cite{Blante96} without
appreciable attention to the phase transitions.

In this paper, we extend the previous work~\cite{ChoiMS98e} on two
capacitively coupled 2D arrays of ultrasmall Josephson junctions to
investigate the charge-frustration effects on the quantum phase transitions.
In a manner similar to that of Ref.~\onlinecite{ChoiMS98f}, we map the system
to the $S=1$ anisotropic Heisenberg antiferromagnet near the particle-hole
symmetry lines and to the $S=1/2$ one near the maximal-frustration lines.
It is then argued that the two spin models can in effect be incorporated into a
single 2D quantum phase model with the effective self-capacitance given by
the coupling capacitance of the original two-array system and the junction
capacitance by the intra-array junction capacitance.  The resulting model
indicates that near the maximal frustration line the system may exhibit a
quantum phase transition from the charge-density wave (CDW) to the super-solid (SS)
phase.  In the SS state, the system possesses
both diagonal and off-diagonal long-range order (DLRO and ODLRO):
Namely, both the density-correlation of charges and the phase-correlation of
superconducting order parameters remain finite as the distance grows
arbitrarily large.\cite{ChoiMS98f,Otterl94}

\begin{figure}
\begin{center}
\epsfig{file=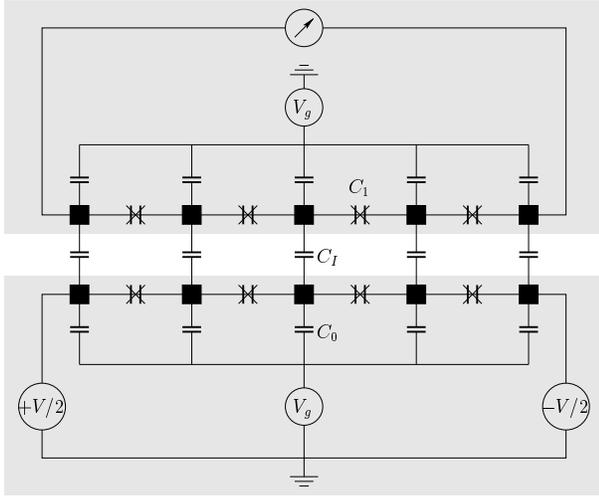,clip=,width=8cm}
\end{center}
\caption{Schematic diagram of the two coupled 2D arrays. Each of the upper
 and lower arrays represents a 2D array composing the system.}
\label{gate:fig1}
\end{figure}

The system of coupled 2D square arrays, shown schematically in Fig.~\ref{gate:fig1},
is described by the Hamiltonian
\beqa \label{gate:1}
 H
 & = & 2e^2\sum_{\ell,\ell';\bfr,\bfr'}[n_\ell(\bfr)-n_g]
     \BfC_{\ell\ell'}^{-1}(\bfr,\bfr')[n_{\ell'}(\bfr')-n_g]
     \nonumber\\&&\mbox{}
   - E_J\sum_{\ell,\bfr,\mu}\cos[\phi_\ell(\bfr)-\phi_\ell(\bfr{+}\hat\bfe_\mu)] 
     \nonumber \\
 & \equiv & H_C + H_J  , 
\eeqa
where the number $n_\ell(\bfr)$ of the Cooper pairs and the phase
$\phi_\ell(\bfr)$ of the superconducting order parameter at site $\bfr$ on the
$\ell$th array ($\ell=1,2$) are quantum-mechanically conjugate variables:
$[n_\ell(\bfr),\phi_\ell(\bfr')]=i\delta_{\ell\ell'}\delta_{\bfr\bfr'}$.
The Josephson coupling between neighboring sites $\bfr$ and $\bfr{+}\hat\bfe_\mu$
(with $\hat\bfe_\mu$ being the unit vector in the direction $\mu = x, y$)
in each array is characterized by the coupling energy $E_J$,
whereas the external charge $n_g\equiv{}C_0V_g/2e$ induced
on each island by the
applied gate voltage $V_g$ breaks the particle-hole symmetry of the system,
introducing charge frustration.
The two arrays are coupled through the capacitance $C_I$ between two grains
at the same position $\bfr$ on the two arrays. (Note the difference from the
Josephson coupled multi-layered system,\cite{Blatte94} where Cooper-pair
tunneling between layers is allowed.) The capacitance matrix
$\BfC$ characterizing the charging energy part $H_C$ of the Hamiltonian in
Eq.~(\ref{gate:1}) can be written in the block form:
\beq\label{gate:2}
 \BfC_{\ell\ell'}(\bfr,\bfr')
 \equiv C(\bfr,\bfr')\left[ \begin{array}{cc}
     1\, & \,0 \\ 0\, & \,1 \end{array} \right]
   + \delta_{\bfr,\bfr'}C_I\left[ \begin{array}{rr}
       1 & -1 \\ -1 & 1 \end{array} \right] ,
\eeq
where $C(\bfr,\bfr')$ is the usual intra-array capacitance matrix 
\beqa\label{gate:3}
 C(\bfr,\bfr')
 & \equiv & C_0\;\delta_{\bfr\bfr'} \nonumber\\
 & + & C_1\sum_{\mu}\left[ 2\delta_{\bfr\bfr'}
      -\delta_{\bfr,\bfr'{+}\hat\bfe_\mu}-\delta_{\bfr,\bfr'{+}\hat\bfe_\mu}
     \right] .
\eeqa
with $C_0$ and $C_1$ being the self- and junction capacitance,
respectively.  Although it is not essential in the subsequent discussion as
long as the interaction range is finite, we assume for simplicity that
${C_1/C_0}\lesssim{1}$, keeping only the on-site and the
nearest-neighbor interactions between the charges.  We also define charging
energy scales $E_0\equiv{e^2/2C_0}$, $E_1\equiv{e^2/2C_1}$, and
$E_I\equiv{e^2/2C_I}$, associated with the corresponding capacitances.

\begin{figure}
\begin{center}
\epsfig{file=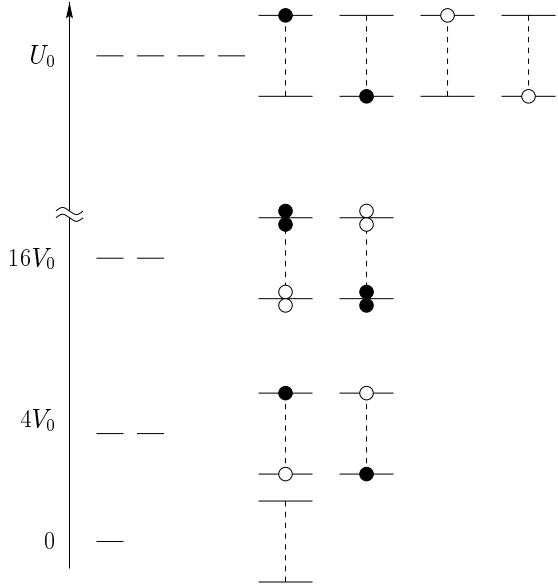,clip=,height=8cm}
\end{center}
\caption{Energy levels of $H_C$ in Eqs.~(\ref{gate:4a}) and 
 corresponding charge configurations near the particle-hole symmetry line.  
 Filled and empty circles denote particles and holes, respectively;
 paired (upper and lower) solid lines represent the two coupled arrays, 
 the couplings between which are illustrated by the dashed lines.
 The low-lying energy levels satisfying $n_+(\bfr)=0$ are well
 separated by a large amount of energy (of the order of $E_0$) from those with
 $n_+(\bfr)\neq0$.}
\label{gate:fig2}
\end{figure}

\begin{figure}
\begin{center}
\epsfig{file=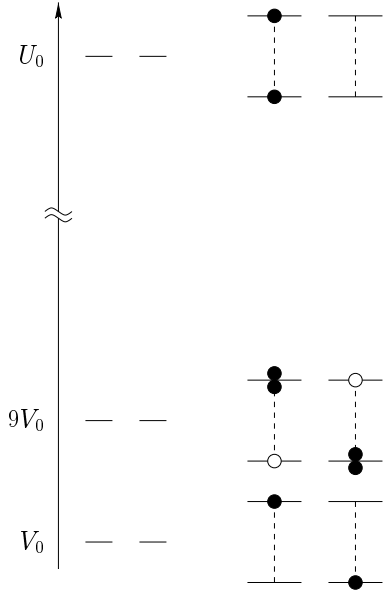,clip=,height=8cm}
\end{center}
\caption{Energy levels and corresponding charge configurations
 near the maximal-frustration line.
 It should be noticed that the ground state is two-fold degenerate per site.}
\label{gate:fig3}
\end{figure}

In the regime of concern in this paper, $C_I\gg{}C_0\;(\gtrsim{}C_1)$, 
i.e., $E_I\ll{}E_0\;(\lesssim{}E_1)$, the charging energy part of the Hamiltonian
in Eq.~(\ref{gate:1}) can be written conveniently as the sum
$H_C=H_C^++H_C^-$ with each component defined to be
\balign
 \label{gate:4a} H_C^+
 & \equiv & U_0\sum_\bfr^{}[n_+(\bfr)-2n_g]^2
   \nonumber\\&&\mbox{}
   + U_1\sum_{\bfr,\mu}[n_+(\bfr)-2n_g][n_+(\bfr{+}\hat\bfe_\mu)-2n_g] \\
 H_C^- & \equiv & V_0\sum_\bfr^{}[n_-(\bfr)]^2
   + V_1\sum_{\bfr,\mu}^{}n_-(\bfr)n_-(\bfr{+}\hat\bfe_\mu) , \nonumber
\ealign
where we have defined new charge variables
$n_\pm(\bfr)\equiv{}n_1(\bfr)\pm{}n_2(\bfr)$ and the interaction
strengths are given by $U_0\simeq{}2E_0$, $V_0\simeq{}E_I$,
$U_1\simeq{}4(C_1/C_0)E_0$, and $V_1\simeq{}(C_1/C_I)E_I$.  This
representation of the charging energy part $H_C$ allows us to
distinguish clearly the two interesting regions from each other:
near the {\em particle-hole symmetry line} $n_g=0$ and near
the {\em maximal-frustration line} $n_g=1/2$, as one can see from the
energy spectra of $H_C$ displayed in Figs.~\ref{gate:fig2} and~\ref{gate:fig3} 
for the two regimes, respectively (recall that $U_0 \gg V_0$).
(Since the system
is invariant with $n_g\to{}n_g+1$, we need to consider only the range
$0\leq{}n_g<1$).
As pointed out for two coupled 1D arrays in Ref.~\onlinecite{ChoiMS98f},
there follow the remarkable properties of the spectrum of $H_C$ in each
regime: Near the maximal-frustration line, the charge configurations which
do not satisfy the condition $n_+(\bfr)=1$ (for all $\bfr$) have a huge
excitation gap of the order of $E_0$.  (Note that we are interested in the
parameter regime $E_I, E_J\ll{}E_0$.) Furthermore, the ground states of
$H_C$, separated from the excited states by the gap of the order of $E_I$,
have two-fold degeneracy for each $\bfr$, corresponding to
$n_-(\bfr)=\pm1$.  This degeneracy is lifted by the Josephson coupling term
$H_J$ in Eq.~(\ref{gate:1}) as $E_J$ is turned on.  As a result, it is
convenient in this case to work within the reduced Hilbert space $\varE_d$,
where $n_+(\bfr)=1$ and $n_-(\bfr)=\pm1$ for each $\bfr$.  Near the
particle-hole symmetry line, on the other hand, the low-energy charge
configuration should satisfy the condition $n_+(\bfr)=0$ for all $\bfr$.
Unlike the former case, the ground state of $H_C$ is non-degenerate and
forms a Mott insulator characterized by $n_1(\bfr)=n_2(\bfr)=0$ for all
$\bfr$.  As $E_J$ is turned on, the ground state of $H_C$ is mixed with the
states with $n_-(\bfr)=\pm2$.  Accordingly, the relevant reduced Hilbert
space is given by $\varE_s$, where $n_+(\bfr)=0$ and $n_-(\bfr)=0,\pm2$ for
all $\bfr$.

\begin{figure}
\begin{center}
\epsfig{file=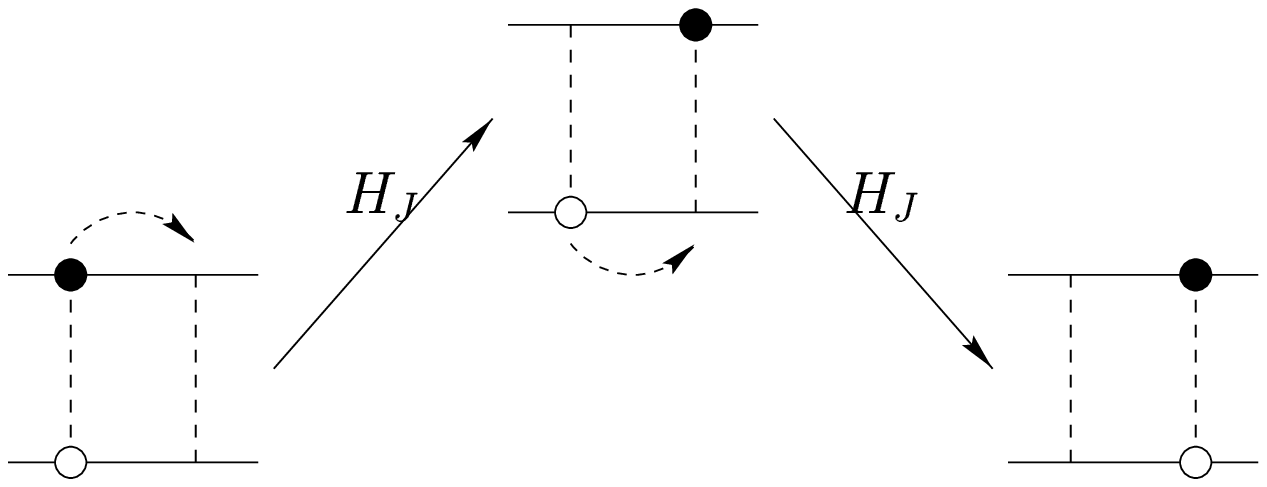,clip=,width=7cm}\ \\(a)\\[5mm]
\epsfig{file=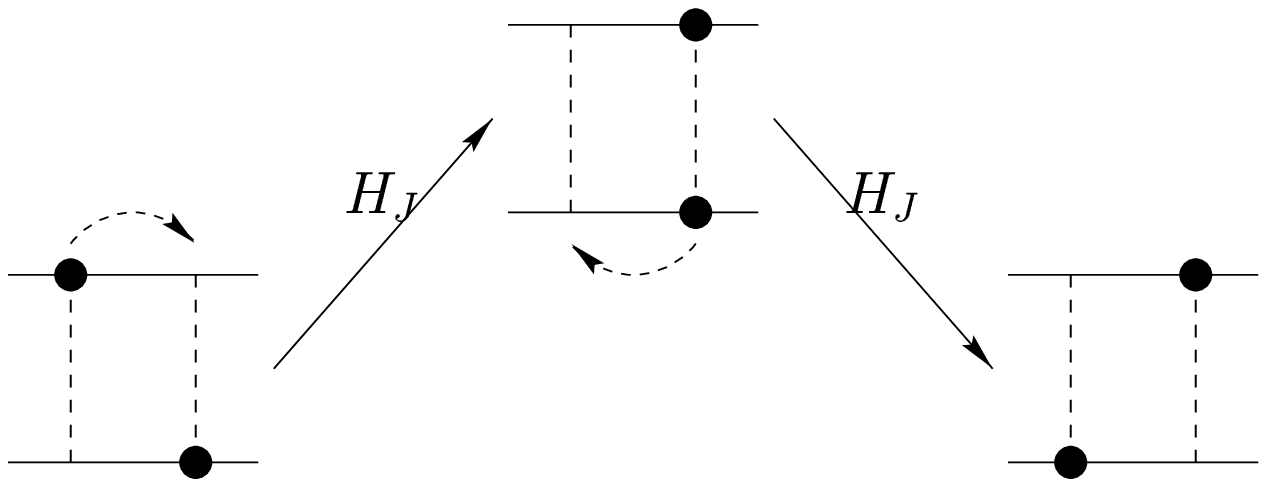,clip=,width=7cm}\ \\(b)
\end{center}
\caption{Typical cotunneling processes relevant (a) near the particle-hole
 symmetry line and (b) near the maximal-frustration line.  The intermediate
 virtual state costs an energy of the order of $E_0$.}
\label{gate:fig4}
\end{figure}

Accordingly, it is instructive to project the Hamiltonian in
Eq.~(\ref{gate:1}) onto $\varE_s$ ($\varE_d$) for $n_g\ll{}1/4$ (for
$|n_g-1/2|\ll{}1/4$); this results in the effective Hamiltonian, up to the
second order in $E_J/E_0$,
\beq\label{gate:5}
 H_{\it eff}
 \equiv P\left[ H + H_J\frac{1-P}{E-H_C}H_J \right]P ,
\eeq
where $P$ is the projection operator.~\cite{Bruder93} 
Explicit implementation of the projection near the particle-hole symmetry
line can be achieved by first noting the correspondence between the
charge picture of the original model and the pseudo-spin ($S=1$) 
picture in the reduced Hilbert space $\varE_s$:
\beqa
 &&S^z(\bfr)
 \equiv P\,\frac{n_1(\bfr){-}n_2(\bfr)}{2}\,P 
 \nonumber \\
 &&S^+(\bfr)
 \equiv \sqrt{2} Pe^{-i\phi_1(\bfr)}(1-P)e^{+i\phi_2(\bfr)}P
 \label{gate:7a} \\
 &&S^-(\bfr)
 \equiv \sqrt{2} Pe^{-i\phi_2(\bfr)}(1-P)e^{+i\phi_1(\bfr)}P.
 \nonumber
\eeqa
In particular, the spin-flip operators $S^+$ and $S^-$ manifest the second-order 
cotunneling process of the particle-hole pairs via an intermediate
virtual state, as depicted in Fig.~\ref{gate:fig4}(a), and mix the energy
levels with unpaired particles or holes of energy $U_0$ and those with
particle-hole pairs of energy $4V_0$ (see Fig.~\ref{gate:fig2}).  It then 
follows that the effective Hamiltonian in Eq.~(\ref{gate:5}) takes the form 
\beqa\label{gate:6}
 &&H_{\it XXZ}^{S=1}
 = \half\gamma_1{J}\sum_\bfr \left[S^z(\bfr)\right]^2 \nonumber\\&&\mbox{}
 - \frac{1}{4}J\sum_{\bfr,\mu}
   \left\{ S^+(\bfr)S^-(\bfr{+}\hat\bfe_\mu) + S^-(\bfr)S^+(\bfr{+}\hat\bfe_\mu)
   \right\} ,
\eeqa
which describes the spin-1 2D XXZ antiferromagnet.\cite{gate:note1} Here
the exchange interaction and the anisotropy ratio are given by
\beq\label{gate:7}
 J \simeq \frac{E_J^2}{4E_0} \quad\mbox{and}\quad
 \gamma_1 \simeq \frac{1}{K^2} ,\quad
\eeq
respectively, with the dimensionless coupling constant
$K\equiv\sqrt{E_J^2/32E_IE_0}$.
Near the maximal-frustration line, on the other hand, the effective
Hamiltonian reduces to that for a spin-1/2 2D XXZ antiferromagnet
\beqa\label{gate:8}
 && H_{\it XXZ}^{S=1/2}
 = \gamma_\half{J}\sum_{\bfr,\mu}
   S^z(\bfr)S^z(\bfr{+}\hat\bfe_\mu) \nonumber\\&&\mbox{}
 - \half{J}\sum_\bfr \sum_\mu
   \left\{ S^+(\bfr)S^-(\bfr{+}\hat\bfe_\mu) + S^-(\bfr)S^+(\bfr{+}\hat\bfe_\mu)
   \right\} ,
\eeqa
with the exchange interaction and the anisotropic ratio given by
\beq\label{gate:9}
 J \simeq \frac{E_J^2}{4E_0} \quad\mbox{and}\quad 
 \gamma_\half \simeq \frac{C_1}{2C_I}\frac{1}{K^2} ,
\eeq
respectively.  In this case, the definitions of the pseudo-spin operators
in terms of the
phase and charge operators are also different slightly from those in Eq.~(\ref{gate:7a}):
\beqa
 &&S^z(\bfr)
 \equiv P\,\frac{n_1(\bfr){-}n_2(\bfr)}{2}\,P \nonumber \\
 &&S^+(\bfr)
 \equiv Pe^{-i\phi_1(\bfr)}(1-P)e^{+i\phi_2(\bfr)}P \label{gate:9a} \\
 &&S^-(\bfr)
 \equiv Pe^{-i\phi_2(\bfr)}(1-P)e^{+i\phi_1(\bfr)}P \nonumber.
\eeqa
Such spin-flip operators are associated with the cotunneling of the
particle-void pairs as displayed in Fig.~\ref{gate:fig4}(b). 

In two dimensions, unlike the 1D case, neither of the two (spin-1 and
spin-1/2)  XXZ antiferromagnets described by Eqs.~(\ref{gate:6}) and (\ref{gate:8})
allow exact solutions.
The simple mean-field theory based on the Ginzburg-Landau
approach~\cite{gate:note2} indicates that the spin-1 XXZ antiferromagnet
may exhibit a zero-temperature phase transition from 
the XY-like phase to the spin-1 Ising-like phase at
$K\sim{}1$ or $\gamma_1\sim{}1$. In the charge picture, the XY-like phase
corresponds to the superconducting (SC) state displaying ODLRO while the
spin-1 Ising-like phase characterized by $S^z(\bfr)=0$ describes the Mott insulator (MI)
state with DLRO.  On the other hand, mean-field-like
approaches~\cite{Bukman92} and numerical approaches~\cite{Barnes89} to
the spin-1/2 XXZ antiferromagnet suggest a zero-temperature phase
transition from the XY-like phase to the spin-1/2 Ising-like phase with
staggered magnetization at $K\sim{}\sqrt{C_1/2C_I}$ or $\gamma_\half\sim{}1$,
corresponding to the SC state and the CDW state, respectively.

Therefore, for the present, the projection of the Hamiltonian to get the
effective spin model does not provide us with direct information about the
phase transitions.  Remarkably, however, the spin models, given by Eqs.~(\ref{gate:6})
and (\ref{gate:8}) in the two regimes, 
can be obtained from a {\em single} 2D quantum phase model (QPM),
via appropriate projections.
This strongly indicates that both regimes can
be described by the Hamiltonian for the QPM:
\beqa
 H_{\it QPM}
 & = & 2e^2\sum_{\bfr,\bfr'}[n(\bfr)-\tilde{n}_g]C^{-1}(\bfr,\bfr')[n(\bfr')-\tilde{n}_g] \nonumber
     \\&&\mbox{}\label{gate:10}
   - \frac{E_J^2}{4E_0}\sum_{\bfr,\mu}
     \cos[\phi(\bfr)-\phi(\bfr{+}\hat\bfe_{\mu})], 
\eeqa
where the effective Josephson-coupling energy $E_J^2/4E_0$ is much
reduced compared with the original intra-array value $E_J$,
and the effective capacitance matrix reads
\beqa
 C(\bfr,\bfr')
 & \equiv & C_I\;\delta_{\bfr\bfr'} \nonumber \\ \label{gate:11}
 & + & \frac{C_1}{2}\sum_{\mu}\left[ 2\delta_{\bfr\bfr'}
      -\delta_{\bfr,\bfr'{+}\hat\bfe_\mu}-\delta_{\bfr,\bfr'{+}\hat\bfe_\mu}
     \right]. 
\eeqa
Note that the self-capacitance is given by $C_I$ instead of the original
value $C_0$ while the junction capacitance is given by $C_1/2$.
The value of charge frustration $\tilde{n}_g$ is related with that ($n_g$)
of the original model given by Eq.~(\ref{gate:1}) in the following way:
At the symmetry line ($n_g=0$) and the maximal-frustration line ($n_g=1/2$)
of the original model, we have $\tilde{n}_g = n_g$.
Near those lines, however, the value of $\tilde{n}_g$ is rather insensitive to
that of $n_g$: 
Namely, $\tilde{n}_g$ remains close to zero and to $1/2$ in the rather large ranges 
around $n_g=0$ and $1/2$, respectively, changing its value sharply 
near $n_g \approx 1/4$.
Accordingly, the QPM in Eq.~(\ref{gate:10})
is either near the symmetry line ($\tilde{n}_g \approx 0$) or near
the maximal-frustration line ($\tilde{n}_g \approx 1/2$)
except for the more or less narrow range around $n_g = 1/4$.

The reduction of the above QPM to the spin-1 and the spin-1/2 XXZ models 
via appropriate projections can be recognized as follows:
In the case $\tilde{n}_g \approx 0$, 
the charging energy reaches its minimum at $n(\bfr)=0$.  
This ground state becomes mixed with the states $n(\bfr)=\pm 1$,
as the Josephson coupling is turned on. 
On the other hand, for $\tilde{n}_g \approx 1/2$, 
the minimum of the charging energy arises at $n(\bfr) = 0, 1$, 
yielding two-fold degenerate ground states.
These situations are essentially the same as those of the original model
with charge frustration $n_g$. 
We thus project
the QPM onto the spaces $\tilde{\varE}_s\equiv\{n(\bfr)=0,\pm1\}$
and $\tilde{\varE}_d\equiv\{n(\bfr)=0,1\}$ 
with the psuedo-spin operators redefined as
  \begin{eqnarray}
    S^z(\bfr) &\equiv& P n(\bfr) P \nonumber \\
    S^+(\bfr) &\equiv& \sqrt{2}P e^{-i \phi(\bfr)}P \\
    S^-(\bfr) &\equiv& \sqrt{2}P e^{i \phi(\bfr)}P \nonumber
  \end{eqnarray}
in space $\tilde{\varE}_s$ and
  \begin{eqnarray}
    S^z(\bfr) &\equiv& P n(\bfr) P -1/2 \nonumber \\
    S^+(\bfr) &\equiv& P e^{-i \phi(\bfr)}P \\
    S^-(\bfr) &\equiv& P e^{i \phi(\bfr)}P \nonumber
  \end{eqnarray}
in space $\tilde{\varE}_d$;
these projections reproduce, in the zeroth order of $E_J/E_I$,
both the spin-1 and the spin-1/2 XXZ
models in Eqs.~(\ref{gate:6}) and (\ref{gate:8})
for $n_g\ll{}1/4$ and $|n_g-1/2|\ll{}1/4$, respectively.

As we proceed to higher orders, the projection
of the single-layer QPM in Eq.~(\ref{gate:10}) in general
yields the coefficients of the $n$th-order terms
$(E_0/E_I)^{(n-1)}$ times larger than those 
in the projection of the original model in Eq.~(\ref{gate:1}).
In spite of such discrepancy in numerical coefficients,
the two projections (of the original model and of the single-layer QPM)
onto their own spin models should bring about quite
similar structures.
For example, mixing of the energy
levels in ${\varE_s}$ with those satisfying $n_-(\bfr)=\pm4,\pm6,\cdots$
(but still keeping $n_+(\bfr)=0$)
always occurs via the virtual states with energies of the order of
$E_0$.  Consequently, at least in the two regimes of concern here,
it is not irrelevant to consider the single-layer QPM in
Eq.~(\ref{gate:10}) as an effective model for the original system.
Quite naturally, the deviation of the QPM in Eq.~(\ref{gate:10}) from the original
model increases with $E_J$.

\begin{figure}
\begin{center}
\epsfig{file=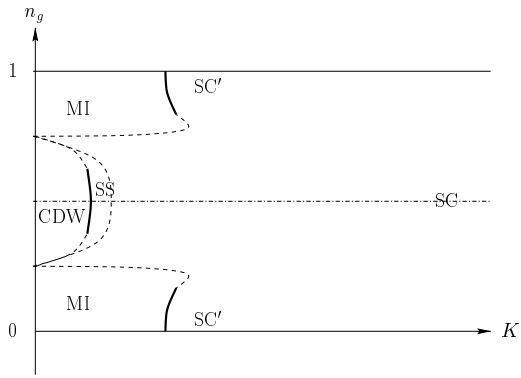,clip=,width=7cm}
\end{center}
\caption{Schematic phase diagram of the system. The superconducting SC$'$
 phase is distinguished from the SC phase in that the underlying transport mechanism
 is the cotunneling process instead of the single-charge transport.}
\label{gate:fig5}
\end{figure}

The 2D QPM has been studied extensively in recent years (see, e.g., Ref.\
\onlinecite{Otterl94}).  Remarkably, for $|n_g-1/2|\ll{}1/4$, it was
suggested that there may exist an unusual SS phase with both the DLRO and
ODLRO, i.e., the coexistence of the crystalline charge ordering together
with superconductivity.  The existence of the SS phase conflicts with the
prediction of a direct transition from the CDW to the SC based on the spin-1/2 XXZ
model in Eq.~(\ref{gate:8}), but such conflict also appears when one simply
truncates the effects of higher energy levels in the QPM.\cite{Otterl94}
These arguments finally yield the schematic phase diagram shown in
Fig.~\ref{gate:fig5}, where the thick solid lines represent the phase boundaries
of the SI transitions, separating the SC from the MI (near the symmetry line) or 
from the CDW (near the maximal-frustration line depicted by the dashed-dotted line). 
Note that these boundaries (near the two lines) change rather gradually as $n_g$ is varied, 
which reflects that near the two lines
the effective charge frustration $\tilde{n}_g$ in the QPM is insensitive to 
the original charge frustration $n_g$. 
The dashed lines in Fig.~\ref{gate:fig5} represent the somewhat speculative
boundaries discussed above;
here the region occupied by the SS phase might be small
because in the QPM the self-capacitance is much larger than 
the junction capacitance ($C_I \gg C_1/2$).\cite{Otterl94}

In conclusion, we have investigated the quantum phase transitions in two
capacitively coupled two-dimensional Josephson-junction arrays
with charge frustration.  The system has been mapped into the $S=1$ and
the $S=1/2$ anisotropic XXZ antiferromagnets near the particle-hole
symmetry line and the maximal-frustration line, respectively.  We have then
argued that the two spin models in effect can be incorporated into a single
quantum phase model.  Based on the resulting model, it has been suggested
that near the maximal frustration line the system may exhibit a quantum phase
transition from the charge-density wave to the super-solid phase, displaying
both diagonal and off-diagonal long-range order.

%
This work was supported in part by the SNU Research Fund, by the Korea Research Foundation,
and by the Korea Science and Engineering Foundation.

%

\begin{thebibliography}{10}

\bibitem[*]{} Present address: Department of Physics and Astronomy,
University of Basel, 4056 Basel, Switzerland.

\bibitem{Averin92}
 D.~V. Averin and Y.~V. Nazarov,  in {\em Single Charge Tunneling: Coulomb
 Blockade Phenomena in Nanostructures}, edited by H. Grabert and M.~H.
 Devoret (Plenum Press, New York, 1992), p.\ 217.

\bibitem{Averin91a}
 D.~V. Averin, A.~N. Korotkov, and Y.~V. Nazarov, Phys. Rev. Lett.
 {\bf 66}, 2818  (1991).

\bibitem{Matter97}
 M. Matters, J.~J. Versluys, and J.~E. Mooij, Phys. Rev. Lett. {\bf 78},
 2469 (1997).

\bibitem{ChoiMS98f}
 M.-S. Choi, M.~Y. Choi, T. Choi, and S.-I. Lee, Phys. Rev. Lett. {\bf 81},
 4240  (1998).

\bibitem{ChoiMS97p7}
 M.-S. Choi, M.~Y. Choi, and S.-I. Lee, preprint (cond-mat/9802237).

\bibitem{ChoiMS98e}
 M.-S. Choi, J. Phys.: Condens. Matter {\bf 10},  9783  (1998).

\bibitem{Blante96}
 {Ya.~M. Blanter} and G. Sch\"on, Phys. Rev. B {\bf 53},  14 534  (1996);
 {Ya.~M. Blanter}, R. Fazio, and G. Sch\"on, Nucl. Phys. B {\bf S58}, 79
 (1997);
 J.~V. Jos\'e, J. Stat. Phys. {\bf 93},  943  (1998).

\bibitem{Otterl94}
 A. van Otterlo and K.-H. Wagenblast, Phys. Rev. Lett. {\bf 72}, 3598
 (1994);
 A. van Otterlo, K.-H. Wagenblast, R. Baltin, C. Bruder, R. Fazio, and G.
 Sch\"on, Phys. Rev. B {\bf 52}, 16176 (1995).

\bibitem{Blatte94}
 G. Blatter, M.~V. Feigel'man, V.~B. Geshkenbein, A.~I. Larkin, and V.~M.
 Vinokur, Rev. Mod. Phys. {\bf 66},  1125  (1994).

\bibitem{Bruder93}
 C. Bruder, R. Fazio, and G. Sch\"on, Phys. Rev. B {\bf 47},  342  (1993);
 L.~I. Glazman and A.~I. Larkin, Phys. Rev. Lett. {\bf 79},  3736  (1997).

\bibitem{gate:note1}
 The sign of the second term in Eq.~(\ref{gate:6}) is unimportant on a
 biparticle lattice; see, e.g., A. Auerbach, {\em Interacting Electrons and
 Quantum Magnetism} (Springer-Verlag, Berlin/Heidelberg, 1994).

\bibitem{gate:note2}
 M.-S. Choi and M.~Y. Choi, unpublished.

\bibitem{Bukman92}
 D.~J. Bukman and J.~M.~J. van Leeuwen, J. Phys. A {\bf 25},  4285  (1992);
 M.-R. Li, Y.-J. Wang, and C.-D. Gong, Z. Phys. B {\bf 102},  129  (1992).

\bibitem{Barnes89}
 T. Barnes, D. Kotchan, and E.S. Swanson, Phys. Rev. B {\bf 39}, 4357 (1989);
 M. Kikuchi, Y. Okabe, and S. Miyashita, J. Phys. Soc. Jpn. {\bf 59}, 492 (1990).

\end{thebibliography}

%
\newpage

\end{document}